\documentclass[twocolumn,prl]{revtex4}
\usepackage{graphics,graphicx}
\usepackage[utf8]{inputenc}
\usepackage{amsmath}
\usepackage{braket}   
\DeclareMathOperator{\sinc}{sinc} 
\usepackage{amsfonts}
\usepackage{amssymb}
\begin{document}
\title{Quantum disordered state of magnetic charges in nanoengineered honeycomb lattice}
\author{G. Yumnam$^{1}$}
\author{Y. Chen$^{1}$}
\author{J. Guo$^{1}$}
\author{J. Keum$^{2}$}
\author{V. Lauter$^{2,*}$}
\author{D. ~K.~Singh$^{1,*}$}
\affiliation{$^{1}$Department of Physics and Astronomy, University of Missouri, Columbia, MO 65211}
\affiliation{$^{2}$Oak Ridge National Laboratory, Oak Ridge, TN 37831}
\affiliation{$^{*}$email: singhdk@missouri.edu, lauterv@ornl.gov}

\begin{abstract}
A quantum magnetic state due to magnetic charges is never observed, even though they are treated as quantum mechanical variable in theoretical calculations. Here, we demonstrate the occurrence of a novel quantum disordered state of magnetic charges in nanoengineered magnetic honeycomb lattice of ultra-small connecting elements. The experimental research, performed using spin resolved neutron scattering, reveals a massively degenerate ground state, comprised of low integer and energetically forbidden high integer magnetic charges, that manifests cooperative paramagnetism at low temperature. The system tends to preserve the degenerate configuration even under large magnetic field application. It exemplifies the robustness of disordered correlation of magnetic charges in 2D honeycomb lattice. The realization of quantum disordered ground state elucidates the dominance of exchange energy, which is enabled due to the nanoscopic magnetic element size in nanoengineered honeycomb. Consequently, an archetypal platform is envisaged to study quantum mechanical phenomena due to emergent magnetic charges.
\end{abstract}

\maketitle
 
The transformation of magnetic moment to magnetic charges\cite{Sondhi} has spurred the exploration of emergent phenomena, such as avalanche of magnetic monopoles, spin solid order of vortex loops and Wigner crystallization of magnetic charges, in two-dimensional artificial magnetic lattice.\cite{Mengotti,Branford,Heyderman3,Rougemaille,Chen} Among the many variants of two-dimensional magnetic lattices,\cite{Rougemaille3,Olson,Gilbert,Nisoli2} a honeycomb structure is of special interest due to its natural manifestation of strong geometric frustration between magnetic moments, replicating the quasi-spin ice configuration.\cite{Nisoli,Tanaka,Qi} The governing Hamiltonian in a magnetic honeycomb lattice consists of three competing energy terms:\cite{Chern,Nisoli} the magnetic dipolar interaction, nearest and next nearest neighbor exchange interactions ($J_1$ and $J_2$, respectively). The dipolar magnetic interaction is comparable to the nearest neighbor exchange energy in a large element size, typically of sub-micrometer length (fabricated using electron-beam lithography technique), honeycomb.\cite{Nisoli} Reducing the element size of magnetic honeycomb reduces the dipolar energy significantly. Thus, a honeycomb made of nanoscopic elements fosters a new energetic regime where exchange interactions, $J_1$-$J_2$ terms, dictate the underlying mechanism behind magnetic ground state. This scenario renders an exciting research venue to explore the occurrence of theoretically formulated, but never observed, quantum disordered magnetic ground state with massive degeneracy. The novel state was originally predicted to arise in two-dimensional bulk material with exchange interaction coupled atomic Ising moments on a honeycomb motif.\cite{Willis,Iqbal,Shastry,Takagi} The one-to-one correspondence between magnetic moment and magnetic charge makes the exploration viable in magnetically frustrated artificial system. The magnetic charges, in essence, are quantum mechanical entities, represented by Pauli matrices.\cite{Oleg,Chern} Yet, they are observed as classical variables in local and macroscopic probes.\cite{Tanaka,Mengotti}

\begin{figure*}
\centering
\includegraphics[width=17.5 cm]{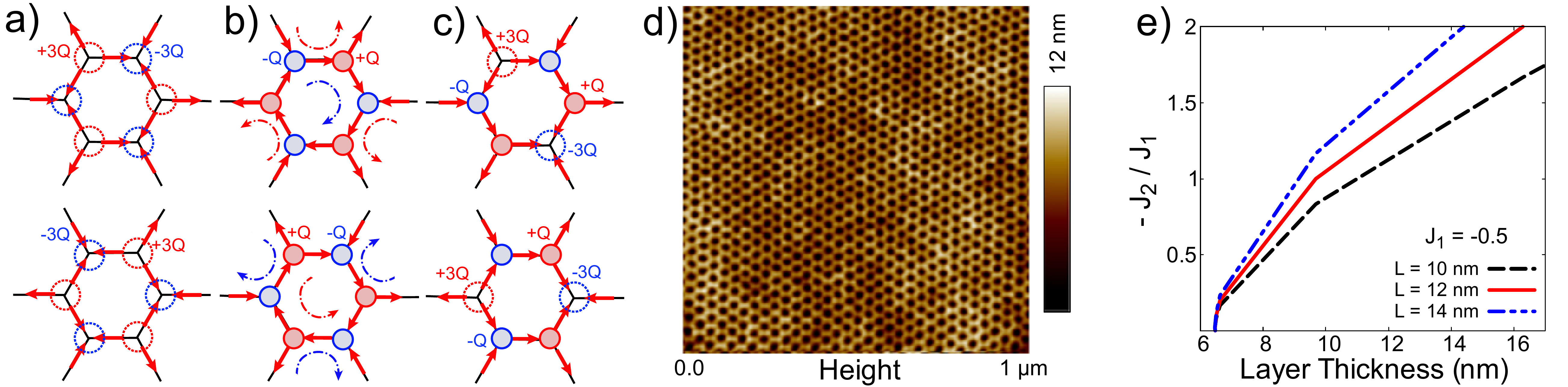} \vspace{-2mm}
\caption{Degenerate arrangements of magnetic charges and variation of exchange constants as functions of geometrical parameters in a honeycomb lattice. (a-c) Schematic description of degenerate configuration of magnetic charges on honeycomb vertices: $\pm$3Q charges in (a), $\pm$Q charges in (b) and $\pm$Q and $\pm$3Q charges in (c). In a large ensemble of vertices, the degeneracy will transform into macroscopic characteristic. Magnetic charge pattern across the lattice transforms by releasing or absorbing a net charge defect of magnitude $\pm$2Q between neighboring vertices. (d) Atomic force micrograph (AFM) of a typical nanoengineered honeycomb lattice, created by the new top down fabrication scheme (see Experimental Section for detail). (e) Plots of $J_2$/$J_1$ as a function of thickness, calculated using Monte-Carlo simulations, for honeycomb element of varying length. $J_1$ and $J_2$ being the nearest neighbor and the next nearest neighbor exchange interactions.} \vspace{-4mm}
\end{figure*}

In this communication, we report the existence of a quantum disordered state of magnetic charges in nanoengineered honeycomb lattice of connecting permalloy (Ni$_{0.81}$Fe$_{0.19}$) elements, with typical dimension of $\simeq$ 12 nm (length)$\times$5 nm (width)$\times$10 nm (thickness). We have employed a hierarchical large throughput nanoengineering technique (described in the Experimental Section) to create an artificial honeycomb lattice of nanoscopic elements. The ultra-small size of nanoscopic element, smaller than the size of a typical permalloy domain $\sim$ 18 nm, paves the way for the dominance of exchange interactions (discussed below) that are necessary to elucidate the pertinent quantum mechanical properties of magnetic charges. This was not possible to achieve in micro-meter element size artificial honeycomb lattice where magnetic dipolar energy is the defining energetic.\cite{Nisoli} Detailed experimental investigation of nanoengineered honeycomb lattice using spin resolved polarized neutron reflectometry measurements has revealed a massively degenerate magnetic charge configurations at low temperature that remain mostly unperturbed to external effect, such as large magnetic field application. The overall magnetization manifests a cooperative paramagnetic behavior as a function of temperature due to charge correlation on honeycomb vertices. The experimental observations in nanoengineered honeycomb e.g. large degeneracy of ground state configuration, independence to external tuning parameter and paramagnetic characteristic are the hallmark of a quantum disordered state.\cite{Willis} The evolution of magnetic charges on honeycomb vertices is key to this novel effect.\cite{Mengotti,Branford} 

There are four possible local moment arrangements on honeycomb vertices: 'two-in \& one-out' (or vice-versa) and 'all-in' or 'all-out' configurations.\cite{Tanaka} The two-in \& one-out arrangement refers to a situation where two moments, aligned along the length of connecting elements due to shape anisotropy, are pointing towards the vertex and one moment is pointing away from it; also termed as the quasi-ice rule.\cite{Qi}The individual moment (treated as Ising spin) can be considered as a pair of '+' and '-' magnetic charges of magnitude Q (directly related to the net moment $\textit{M}$ under the dumbbell representation by Q = $\textit{M}$/$\textit{l}$ where $\textit{l}$ being the length of the connecting element) that interact via magnetic Coulomb's interaction.\cite{Branford,Sondhi,Moessner} Consequently, a 'two-in \& one-out' (or vice-versa) moment arrangement imparts the net charge of +Q (-Q) to a given vertex, whereas the 'all-in or all-out' configurations give rise to more energetic $\pm$3Q charges.\cite{Tanaka,Nisoli} Recently, it was demonstrated that the domain walls in a sufficiently large vertex in an artificial spin ice can take various shapes or, can be made of various sizes.\cite{Perrin} However, the nanoscopic characteristic of our honeycomb lattice forbids multi-domain structure of a given vertex. Magnetic charges on neighboring vertices can emit or absorb a net charge defect, equal to the net difference of $\pm$2Q charge, under external tuning parameters of applied magnetic field or temperature variation in a thermally tunable system.\cite{Mengotti,Branford,Artur,Sendetskyi} Additionally, the charges can arrange themselves in multiple different configurations across the honeycomb vertices without affecting the overall energy of the system, schematically described in Fig. 1a-c.\cite{Oleg,Moessner} For a macroscopic ensemble of the charge hosting vertices, it causes very large degeneracy in the system.

The exchange interaction terms become dominant as the size of the constituting elements of a magnetic honeycomb lattice reduces to the nanoscopic level. Unlike the very large magnetic dipolar interaction energy, $\simeq$10$^{4}$ K, between micrometer size elements,\cite{Nisoli} the strength of dipolar interaction is significantly smaller, $\simeq$ 10 K, in our nanoengineered honeycomb lattice. The reduced inter-elemental dipolar energy increases the occurrence probability of $\pm$ 3Q charges at low temperature; typically forbidden under the dipolar interaction model. In the case of artificial square lattice of permalloy elements, Monte-Carlo (MC) simulations have shown that $J_2$ can be comparable to $J_1$ in moderately thick sample.\cite{Moessner,Moessner2} Therefore, the next nearest neighbor exchange interaction can be the second competing term in the Hamiltonian (besides the nearest neighbor exchange $J_1$), describing the phase transition process in a nanoengineered magnetic lattice. Accordingly, the governing Hamiltonian of the system can be written as:\cite{Chern}

\begin{eqnarray}
{H}&=&-\frac{J_1}{2}{\sum_{nn}{\sigma_i}{\sigma_j}} - \frac{J_2}{2}{\sum_{nnn}{\sigma_i}{\sigma_j}}
\end{eqnarray}

where $J_1$ $<$ 0, $J_2$ $>$ 0 indicating antiferromagnetic and ferromagnetic interactions, respectively. The nature of the second term is chosen to be the same as the dipolar term, as employed in the large element size honeycomb. Here, $\sigma$$_{i}$ and $\sigma$$_{j}$ are quantum Pauli matrices of Ising variable ($\textbf{S}$$_{i}$ = $\sigma_i$$\textbf{e}$$_{i}$ and $\textbf{e}$$_{i}$.$\textbf{e}$$_{j}$ = 1/2) on neighboring vertices, related to the net magnetic charges by $Q_\alpha$ = $\pm$$\sum$$\sigma_i$ where $\sigma_i$ = $\pm$1.\cite{Sondhi,Chern,Oleg} We have performed detailed Monte-Carlo simulation to analyze the competing nature of exchange constants. The simulations were performed for different initialization parameters of exchange constant $J_1$ (see Experimental Section for detail). In Fig. 1e, we plot $J_2$/$J_1$ as a function of thickness for different element size of the honeycomb lattice. For the initialization value of $J_1$ between 0.13 - 0.6 (previously used for numerical simulation to determine the validity of $J_1$-$J_2$ model in geometrically frustrated lattice),\cite{Iqbal} comparable strengths of $J_1$ and $J_2$ are inferred to arise in the moderately thick, $\simeq$ 10 nm, honeycomb lattice (also see Fig. S1-S2 in Supporting Information). Thus, our nanoscopic honeycomb provides an ideal setup to realize the $J_1$-$J_2$ model, which can lead to a macroscopically degenerate ground state of magnetic charges with disordered paramagnetic configuration.\cite{Willis,Iqbal,Shastry} In the following paragraphs, we discuss experimental results to this effect.

\begin{figure*}
\centering
\includegraphics[width=18 cm]{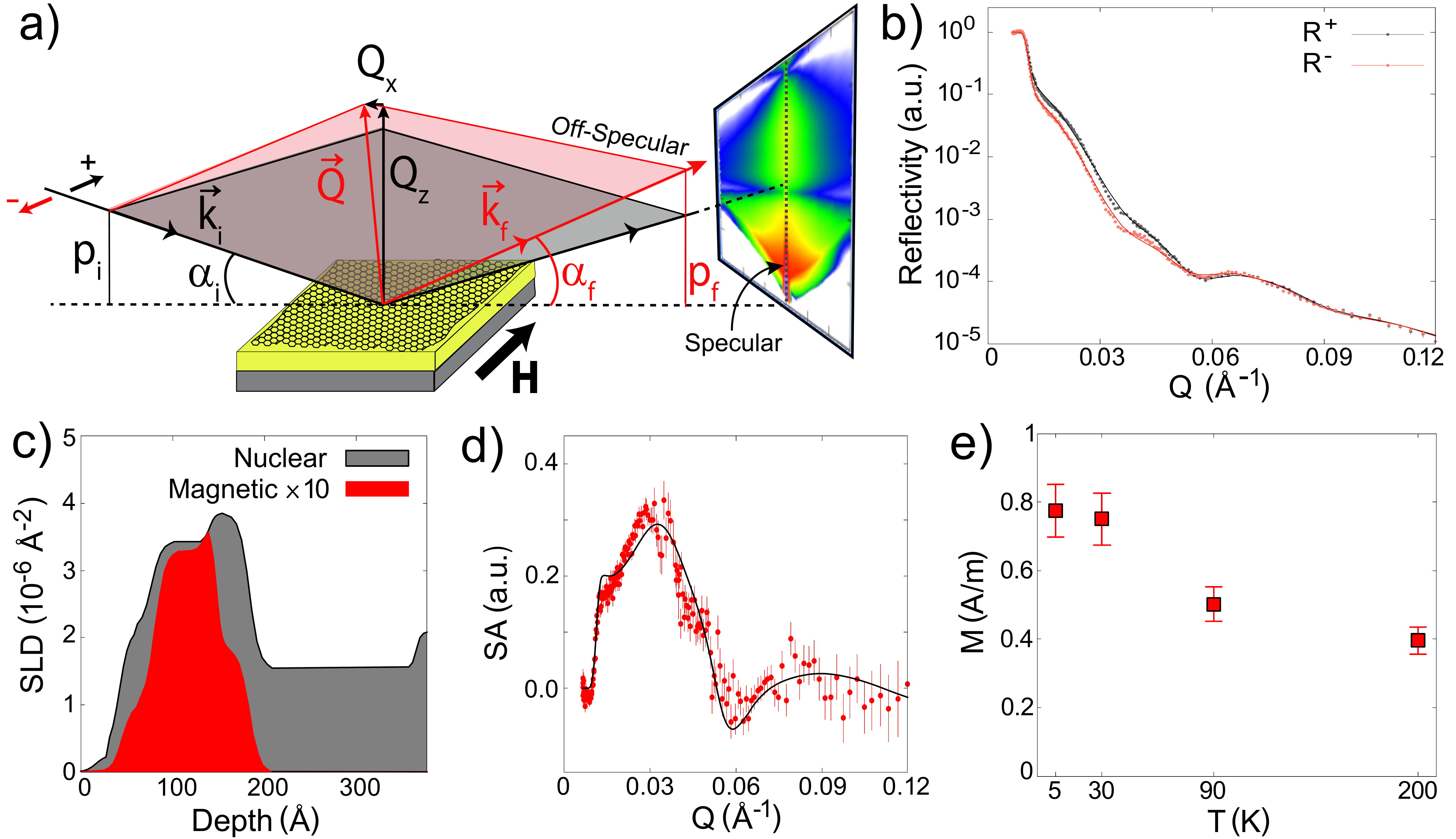} \vspace{-4mm}
\caption{Spin resolved neutron scattering measurements of nanoengineered honeycomb. (a) Schematic design of grazing incidence polarized neutron scattering (PNR) experiment on an artificial magnetic honeycomb lattice. A small guide field of $H$ = 20 Oe is applied to keep neutron polarized. The schematic description depicts the scattering profiles of neutrons, leading to a 2D pattern of specular and off-specular intensities. (b) Measured and fitted (solid curves) reflectivity curves for neutron with spin-up (R$^{+}$) and spin-down (R$^{-}$) polarizations as a function of wave vector transfer at a characteristic temperature of $T$ = 5 K (see Fig. S5 for other temperatures). (c) Nuclear and magnetic scattering length density profiles, obtained from the fit to experimental data, as a function of depth of honeycomb element. (d) Plot of spin asymmetry SA at $T$ =  5 K, obtained from the experimental and fitted reflectivity in fig. c. (e) Estimated net magnetization from PNR measurements as a function of temperature, revealing weak paramagnetic-type correlation in permalloy honeycomb. The lack of divergence in magnetization data indicates the absence of long range order in the system. Error bar in experimental data, in Figs b, d and e, represents one standard deviation. PNR measurements were repeated on two different samples, fabricated under identical conditions. Experimental results were found to be similar.
} \vspace{-4mm}
\end{figure*}

We have performed detailed neutron scattering measurements on $\simeq$1-sq. inch size nanoengineered honeycomb sample, depicted in Fig. 1d, on the Magnetism Reflectometer at BL-4A beamline at the Spallation Neutron Source of Oak Ridge National Laboratory. To understand the nature of lateral magnetic correlation in honeycomb system, we have employed the highly versatile depth-sensitive technique of  grazing incidence polarized neutron reflectometry (PNR) with off-specular scattering.\cite{Peter1,Valeria2}  PNR measurements were performed in a small guide field of $H$ = 20 Oe to maintain the polarization of incident and scattered neutron (see Experimental Section for detail). Schematic of the measurement procedure is shown in Fig. 2a. Experimental data were collected at multiple temperatures between $T$= 300 K and $T$ = 5 K in zero and saturating magnetic field of $H$ = 0.5 T, applied in-plane to the sample. Both spin polarized specular reflectivity and off-specular two dimensional (2D)  scattering patterns were obtained at each temperature and field. We show the plots of specular reflectivities R$^{+}$ and R$^{-}$ ('+' and '-' correspond to neutron with spin parallel and anti-parallel to guide magnetic field) at a characteristic temperature of $T$ = 5 K in Fig. 2b (see Fig. S5 for other temperatures). Two features are clearly observed in the plots: (a) the reflectivities R$^{+}$ and R$^{-}$ are separated from each other, indicating the presence of net magnetic moment in the system, and (b) the splitting between R$^{+}$ and R$^{-}$ gradually increases as temperature reduces to $T$ = 5 K (see Fig. S5). The quantitative determination of the change in the overall magnetization at different temperatures is obtained by analyzing the experimental results using a generic LICORNE-PY program, which generates the reflectivity pattern for a given set of physical parameters of the system e.g. layer thickness, density, interface roughness and magnetic moment of magnetic layer. The fitting was performed simultaneously to the data sets measured at different temperatures and fields, including the X-ray reflectivity (XRR, Fig. S4). This procedure ensured the high accuracy of obtained parameters. Nuclear and magnetic scattering length densities (NSLD and MSLD), corresponding to the depth profiles of chemical structure and in-plane magnetization vector distributions, respectively, are depicted in Fig. 2c. The magnetization of permalloy honeycomb lattice is manifested by the spin asymmetry SA, given by (R$^{+}$ - R$^{-}$)/( R$^{+}$ + R$^{-}$). Experimental plot of spin asymmetry SA as a function of wave vector transfer Q at $T$ = 5 K is shown in Fig. 2d (see Fig. S5 for other temperatures). Estimated magnetization at different temperatures is shown in Fig. 2e.  Expectedly, the net magnetization increases when temperature is reduced. But there is no divergence at low temperature, which suggests the absence of long range order in the system. Additionally, the magnetization does not change much between an intermediate temperature of $T$ = 30 K and $T$ = 5 K.The net magnetization nearly doubles in magnitude between $T$ = 200 K and $T$ = 5 K. Overall, such a trend is indicative of the cooperative paramagnetic behavior.\cite{Gardner2} The paramagnetic behavior is independently verified using bulk magnetization measurement as a function of temperature at low field, see Fig. S3. It is arising due to an increase in the population density of $\pm$Q charges at lower temperature. At high temperature, the honeycomb vertices are occupied by both $\pm$Q and $\pm$3Q charges. As temperature reduces, the population of energetic $\pm$3Q charges decreases. We note that $\pm$Q charges give rise to the finite magnetization due to the short-range correlation of '2-in \& 1-out' (or vice-versa) moment arrangement. On the other hand, vertices with $\pm$3Q charges have zero net magnetization. Hence, there are two possibilities behind the paramagnetic behavior: a, the energetic high integer charges are still present at low temperature with significant population density i.e. not all vertices contribute to the net magnetization as if it is a fragmented state or b, the honeycomb vertices are occupied by only +Q and -Q charges at low temperature and the system has a preference for one of the polarities. We have employed the analysis of off-specular PNR data to understand the underlying mechanism.

\begin{figure*}
\centering
\includegraphics[width=16.5 cm]{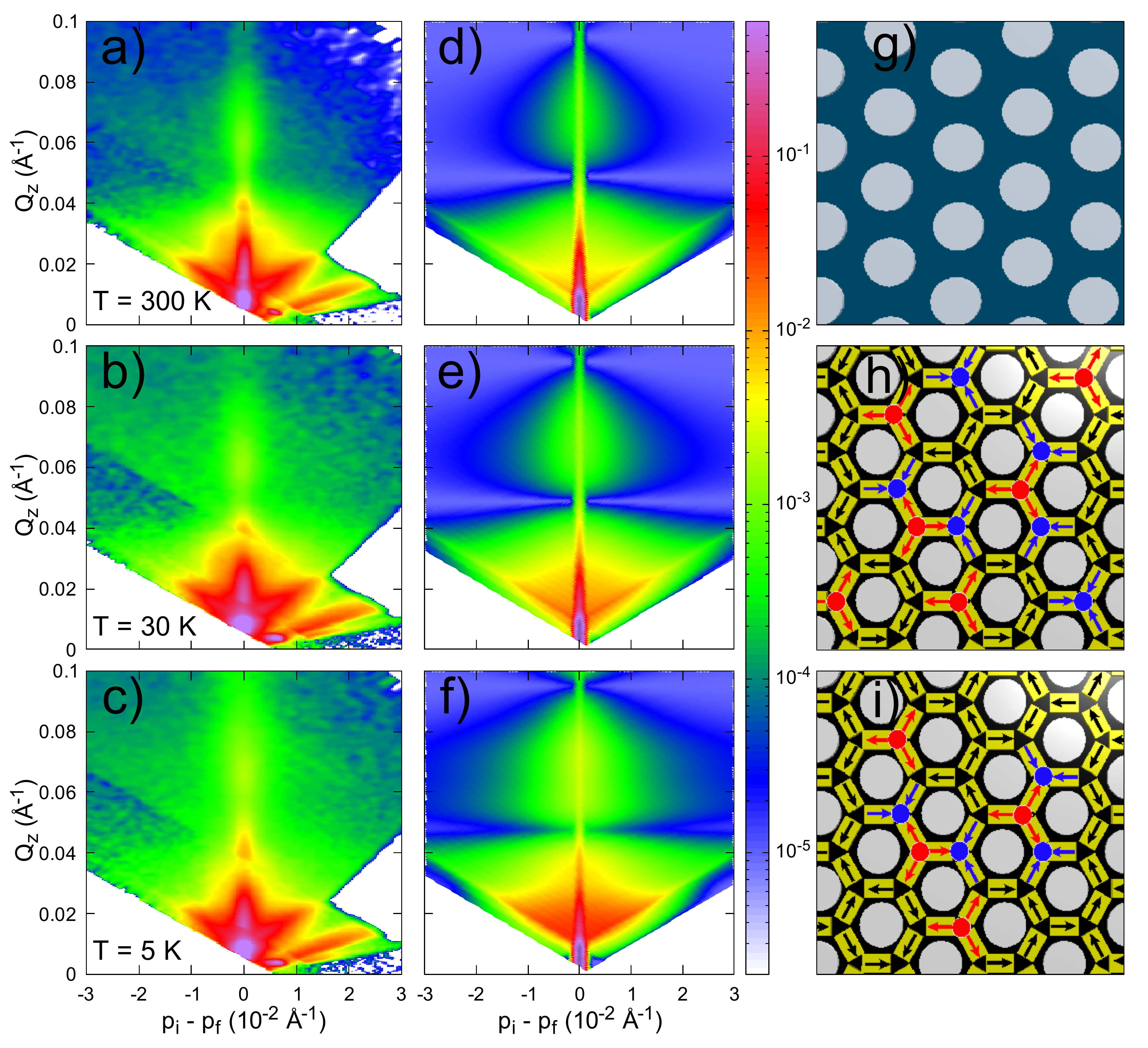} \vspace{-4mm}
\caption{Evolution of quantum disordered state in nanoengineered permalloy honeycomb lattice. (a-c) Experimental two-dimensional (2D) maps of neutron intensity as a function of $(k_i - k_f)$ at characteristic temperatures. Here $k_i$ and $k_f$ are the perpendicular components of incoming and outgoing neutron, respectively (see schematic in Fig. 2a. The 2D maps are obtained by summing both (+ -) and (- +) scattering components. With decreasing temperature, diffuse scattering in the off-specular channel becomes stronger. (d-f) Numerically simulated 2D maps for the corresponding charge configurations shown in Fig. (g-i). Numerical calculation of magnetic structure factor is performed using the DWBA formalism (see text). The highly degenerate disordered charge correlation, involving both $\pm$Q and $\pm$3Q, well describes the experimental data at intermediate ($T$ = 30 K) and low temperature (also see Fig. S6). Blue and red balls indicate the presence of +3Q and -3Q charges, respectively.
} \vspace{-4mm}
\end{figure*}

Off-specular PNR results at few characteristic temperatures are shown in Fig. 3a-c. Here, y-axis represents the out-of-plane scattering vector ($Q_z= \frac{2\pi}{\lambda} (\sin{\alpha_i} + \sin{\alpha_f})$), while the difference between the z-components of the incident and the outgoing wave vectors ($p_i - p_f$ = $\frac{2\pi}{\lambda}(\sin{\alpha_i} - \sin{\alpha_f})$) is drawn along the x-axis (as shown schematically in Fig. 2a). Thus, vertical and horizontal directions correspond to the out-of-plane and in-plane correlations, respectively. \cite{Valeria,Artur}. The specular reflectivity lies along the x = 0 line. At $T$ = 300 K, very small off-specular scattering is detected, which could be arising due to the ferromagnetic nature of permalloy film and the structure of honeycomb lattice. Most of the scattering is confined to the specular line. As temperature is reduced, magnetic diffuse scattering starts developing along the horizontal direction. The broad diffuse scattering in the off-specular data at intermediate temperature of $T$ = 30 K manifests significant in-plane correlation of magnetic moments along honeycomb elements. At further reduced temperature of $T$ = 5 K, the diffuse scattering becomes more prominent. We also observe a peculiar inverted cone-shape in the off-specular PNR data at intermediate temperature between $Q_z$ = 0.09 and 0.05 $\AA$$^{-1}$, which becomes stronger as temperature decreases. Numerical modeling of experimental data is performed using the Distorted Wave Born Approximation (DWBA), as utilized in the BornAgain \cite{bornagain}platform (see Experimental Section for detail). Using DWBA simulations, we generate reflectometry patterns of both specular and off-specular reflectivity. A thin layer of air condensation on the surface of the sample, typical in low temperature PNR measurement, is accounted for in the numerical calculation.\cite{Steffens} As we can see in Fig.3d, the experimental data at $T$ = 300 K is well-described by the scattering from the honeycomb structure. When measurement temperature is reduced to $T$ = 30 K, the numerically simulated reflectometry pattern, Fig. 3e, for the magnetic charge configuration, comprised of both $\pm$Q and $\pm$3Q charges as shown in Fig 3h, is found to be in good agreement with experimental result. Numerical simulations suggest that a significant number of honeycomb vertices are occupied by the $\pm$3Q charges. The short-range magnetic moment correlation, associated to $\pm$Q charges, contribute to the overall magnetization, as shown in Fig. 2e. Most interestingly, the magnetic charge configuration manifests massive degeneracy. Both $\pm$Q and $\pm$3Q charges can be arranged in the multitude of possibilities without affecting the overall energy of the system. The degenerate charge configurations give rise to the exactly same reflectometry pattern (see Fig. S6). At the lowest measurement temperature of $T$ = 5 K, the simulated pattern due to $\pm$Q and $\pm$3Q charges reproduces the shape of off-specular scattering data. Besides the shape and intensity of off-specular scattering profile, the reflectivity below Yoneda lines (at low Q) is reproduced reasonably well in the simulated plot. The horizontal width of simulated profile at $T$ = 5 K appears slightly larger than the experimental data, as the actual sample is much larger in size than the simulated honeycomb lattice. The population density of $\pm$Q charges is enhanced at the expense of $\pm$3Q charges, yet occur in large numbers. Consequently, the overall magnetization increases by $\simeq$ 10\% between $T$ = 30 K to $T$ = 5 K. For completeness, we have also performed numerical modeling of the theoretically predicted pure states, such as spin ice and long range spin solid states (see Fig. S7). Compared to the pure states, mixed phases of $\pm$Q and $\pm$3Q charges manifest distinct reflectometry profiles that are in congruence with experimental data. 

There are two important findings that need to be highlighted here: (a) $\pm$3Q charges persist to the lowest temperature in the lattice. It is a major departure from the prevailing understanding that the high integer charges cannot occur at low temperature in artificially frustrated lattice.\cite{Nisoli} This becomes possible due to the nanoscopic element size in nanoengineered honeycomb, which reduces the inter-elemental energy to $\simeq$ 10 K. (b) Numerical modeling does not indicate any large change in the magnetic charge configuration pattern at low temperature, compared to the intermediate temperature, in the honeycomb lattice. We argue that the massive degeneracy in magnetic charge configuration forbids the development of a stable ground state. The presence of $\pm$3Q charges is pertinent to the occurrence of the disordered state with macroscopic degenerate characteristic. Also, the macroscopic degeneracy of magnetic charges persists at low temperature, in the absence of thermal fluctuation. It elucidates the intrinsic nature of the phenomena. The persistence of macroscopic degeneracy to low temperature highlights the quantum mechanical nature of magnetic charges. Hence, the system manifests a quantum disordered magnetic state at low temperature, with an overall paramagnetic moment arising due to the very short-range correlation between magnetic moments associated to $\pm$Q charges on the non-correlated vertices. This observation is in good agreement with the theoretical prediction of a disordered state in $J_1$-$J_2$ Ising model on the honeycomb lattice.\cite{Willis} Here, it is worth mentioning the effect of disorder on magnetic properties in artificial spin ice. In $\mu$m element size honeycomb lattice, magnetic domains are often pinned by inhomogeneity or disorder in the lattice.\cite{Qi} In some cases, disorder is used as a tuning agent to explore novel properties. For instance, a recent research work by Saccone et al. has shown that the artificially induced disorder of Gaussian form in lattice construction can be used to tune magnetic properties between ferromagnetic and spin glass type behavior in arrays of Ising nanomagnets.\cite{Saccone} A modest thickness fluctuation, to the tune of 1-2 nm (see Fig. S9), is observed across the plaquette in our honeycomb sample. Yet, the overall thickness is smaller than the size of a typical permalloy domain, $\sim$ 18 nm. Hence, the effect of disorder due to modest thickness variation in experimental observation can be ruled out in this case.

\begin{figure}
\centering
\includegraphics[width=8.8 cm]{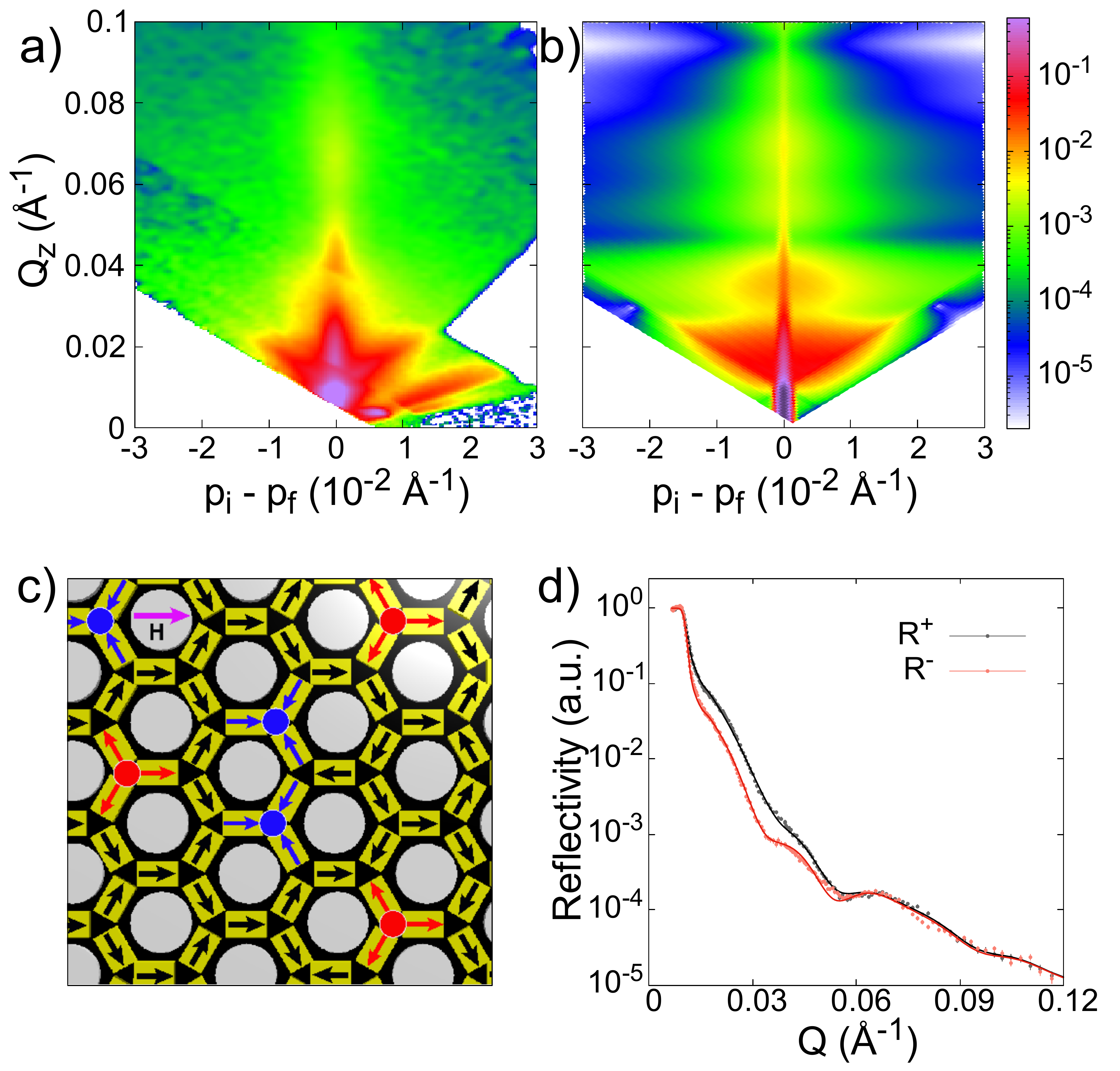} \vspace{-6mm}
\caption{Magnetic field effect on degenerate charge configuration. (a) In applied magnetic field of $H$ = 0.5 T, larger than the coercivity of the permalloy honeycomb lattice ($\simeq$ 0.1 T), the overall magnetism at $T$ = 5 K, as manifested by the specular reflection (Fig.d), is nominally stronger than zero field (magnetic field is applied inplane to the sample). The diffuse scattering in the off-specular data remains mostly unaffected to field application. It is also confirmed by numerically simulated pattern, Fig. b, for the degenerate charge correlation shown in Fig. c. Field application does not lift the degeneracy. (See Fig. S8 for spin asymmetry plot). Error bar in Fig. d represents one standard deviation.
} \vspace{-6mm}
\end{figure}

A quantum disordered magnetic state would reject external influences, such as magnetic field application, that can reduce the degrees of freedom. We test the magnetic field effect on the prevailing disordered state of fragmented $\pm$Q and $\pm$3Q charges at low temperature. Additional measurements were performed in an inplane applied magnetic field of $H$ = 0.5 T, which is significantly larger than the coercivity ($H_c$ $\simeq$ 0.1 T) of magnetic honeycomb lattice.\cite{Summers} The polarized reflectometry pattern in applied field is shown in Fig. 4a. Field application of this magnitude exerts strong enough torque to align Ising moments along the field direction. Consequently, the splitting between R$^{+}$ and R$^{-}$ becomes nominally stronger, as shown in Fig. 4d, compared to $H$ = 0 T. However, the broad feature in diffuse scattering and its inverted cone shape remain intact. Numerical modeling of experimental PNR pattern takes into account the field effect on Ising moments. The best fit to experimental data is obtained for the degenerate charge configuration with higher density of $\pm$Q charges (see Fig. 4b-c). It explains the enhanced splitting between R$^{+}$ and R$^{-}$ in specular reflectivity pattern in Fig. 4d (see Fig. S8 in Supporting Information for spin asymmetry and MSLD plots). The numerical simulation for the aforementioned magnetic charge configuration, Fig. 4c, reproduces the shape and the intensity of the diffuse scattering. Magnetic field application does not lift the degeneracy, albeit nominally reduces it. The system defies external influences and still maintains a very high level of degeneracy. This suggests that the disordered state of mixed charges is indeed robust.

Two-dimensional nanoengineered magnetic system with geometrically frustrated motif not only provides an alternative platform to study the novel and emergent magnetic properties of bulk materials, but also allows us to explore new phenomena in the reduced degrees of freedom. In this communication, we have presented experimental evidences to the occurrence of quantum disordered state of magnetic charges in artificial magnetic honeycomb lattice of nanoscopic element, smaller than the size of a permalloy domain.\cite{Coey} The disordered state, comprised of both $\pm$Q or $\pm$3Q charges, is characterized by macroscopic degeneracy, which persists to low temperature. Typically, the high multiplicity charges, $\pm$3Q, are not stable due to higher energy. Therefore, the lattice is constantly seeking a low energy state by releasing or absorbing $\pm$2Q charge defects between vertices. Yet, a large number of vertices are still occupied by $\pm$3Q charges at low temperature, even in applied magnetic field. In principle, magnetic field application to the tune of $H$ = 0.5 T is capable to align moments to field direction. Consequently, the lattice will be free of high multiplicity charges. This does not seem to be the case here. It further suggests that the observed disorder, due to degenerate distribution of $\pm$Q and $\pm$3Q charges, is intrinsic in nature. The persistent disorderness to the lowest measurement temperature points to the quantum mechanical nature of magnetic charge quasi-particles. After all, magnetic charges are related to the quantum Pauli matrices via the charge sum rule under the dumbbell model.\cite{Sondhi,Chern} Our observation is also consistent with a previous theoretical study, which predicts the occurrence of a quantum disordered state in two-dimensional Ising system.\cite{Willis,Takagi} According to the theoretical researches, an Ising Kagome system with antiferromagnetic nearest neighbor and ferromagnetic next nearest neighbor interactions is expected to depict three key properties: a disordered ground state with massive degeneracy where strong fluctuation nullify the development of any long range order, the persistence of disordered state from moderate to low temperature and magnetic field independence of the ground state degeneracy. All three premises of the theoretical prediction are fulfilled in two-dimensional nanoengineered honeycomb lattice of connected permalloy elements. The presence of disordered ground state in an artificial honeycomb lattice elucidates the quantum mechanical properties of magnetic charges. Future experimental research works on the investigation of dynamic properties of magnetic charges using microwave excitation or optical pumping method are highly desirable to further understand the energetic profile of  the quantum state of magnetic charges in permalloy honeycomb lattice. 

The new honeycomb lattice design has spurred the exploration of emergent magnetic phenomena that are at the forefront of research. Besides the observation of quantum disordered state in moderately thick lattice, previous research works on thinner analogue of artificial honeycomb have demonstrated the development of spin solid type loop state at low temperature.\cite{Artur,Summers} Unlike the collective paramagnetic behavior in magnetization data at low temperature in the disordered state, loop state tends to achieve a near zero magnetization as T$\rightarrow$ 0 K. The opposing magnetization behaviors as a function of varying thicknesses hint of a comprehensive research platform in artificial magnetic honeycomb lattice of nanoscopic elements.

\section{Experimental Section}

$\textit{Sample Fabrication and Characterizations}$: Fabrication of artificial honeycomb lattice involves the synthesis of porous hexagonal diblock template on top of a silicon substrate, calibrated reactive ion etching (RIE) using CF$_{4}$ gas to transfer the hexagonal pattern to the underlying silicon substrate and the deposition of magnetic material (permalloy) on top of the uniformly rotating substrate in near-parallel configuration ($\simeq 1^{o}$) to achieve the two-dimensional character of the system. The sample fabrication process utilized diblock copolymer polystyrene(PS)-b-poly-4-vinyl pyridine (P4VP) of molecular weight 29K Dalton with the volume fraction of 70\% PS and 30\% P4VP. The diblock copolymer tends to self-assemble, under right condition, in a hexagonal cylindrical structure of P4VP in the matrix of polystyrene (PS).\cite{Russell} A 0.5\% PS-b-P4VP copolymer solution in toluene was spin casted onto cleaned silicon wafers at 2500 rpm for 30 s and placed in vacuum for 12 hours to dry. The samples were solvent annealed at 25$^{o}$ C for 12 hours in a mixture of THF/toluene (80:20 v/v) environment. The process results in the self-assembly of P4VP cylinders in a hexagonal pattern within a PS matrix.  The average diameter of a P4VP cylinder is $\simeq$ 15 nm and the center-to-center distance between two cylinders is $\simeq$ 30 nm, also consistent with that reported by Park et al.\cite{Russell}. Submerging the samples in ethanol for 20 minutes releases the P4VP cylinders yielding a porous hexagonal template. The diblock template is used as a mask to transfer the topographical pattern to the underlying silicon substrate. The top surface of the reactively etched silicon substrate resembles a honeycomb lattice pattern. This property is exploited to create metallic honeycomb lattice by depositing permalloy, Ni$_{0.81}$Fe$_{0.19}$, in near parallel configuration in an electron-beam evaporation. For this purpose, a new sample holder was designed and setup inside the e-beam chamber. The substrate was rotated uniformly about its axis during the deposition to create uniformity. This allowed evaporated permalloy to coat the top surface of the honeycomb only, producing the desired magnetic honeycomb lattice with a typical element size of 12 nm (length)$\times$~5 nm (width)$\times$10 nm (thickness). Atomic force micrograph of a typical honeycomb lattice is shown in Fig. 1a. The center-to-center spacing between neighboring honeycombs is $\simeq$ 30 nm. Thus, each honeycomb is about 30 nm wide. 

$\textit{Neutron Scattering Measurements}$: Neutron scattering measurements were performed on a 20$\times$20 mm$^{2}$ surface area sample at the Magnetism Reflectometer, beam line BL-4A of the Spallation Neutron Source (SNS), at Oak Ridge National Laboratory. Neutron measurements were also reproduced on another permalloy honeycomb sample with similar geometrical dimensions, nanofabricated under the identical conditions. The instrument utilizes the time of flight technique in a horizontal scattering geometry with a bandwidth of 5.6 $\AA$(wavelength varying between 2.6 - 8.2 $\AA$). The beam was collimated using a set of slits before the sample and measured with a 2D position sensitive $^{3}$He detector with 1.5 mm resolution at 2.5 m from the sample. The sample was mounted on the copper cold finger of a close cycle refrigerator with a base temperature of $T=5$ K. Beam polarization and polarization analysis was performed using reflective super-mirror devices, achieving better than 98\% polarization efficiency over the full wavelength band. For reflectivity and off-specular scattering the full vertical divergence was used for maximum intensity and a 5\% $\Delta$$\theta$/$\theta$ $\simeq$ $\Delta$$q_z$/$q_z$ relative resolution in horizontal direction. 

\textit{Monte-Carlo simulation}: We have performed Monte-Carlo simulation based on the loop algorithms~\cite{todo}, as implemented in ALPS~\cite{alps}. We have employed the Heisenberg  exchange Hamiltonian: $\mathbf{H} = -J_1 \sum_{\braket{i,j}}\mathbf{\sigma}_i\mathbf{\sigma}_j - J_2 \sum_{\braket{i',j'}}\mathbf{\sigma}_i\mathbf{\sigma}_j$, where $\mathbf{\sigma}_i$ is the Pauli spin-1/2 operator at the lattice vertex sites $i$ (see the main text), $J_1$, $J_2$ are nearest and next-nearest neighbor exchange interaction parameters. For numerical simulations, we have used a kagome lattice as the host sites for $\sigma$, which arises due to the moment aligned along the length of honeycomb element. The near-zero temperature properties of the spin-system was calculated at sufficiently low temperature, or large $\beta = 1/K_BT$ such that we can consider our system as a converged ground-state with $5\times 10^4$ sweeps per thermalization step. The ground state energy $E_0$ calculation is repeated for different values of $J_1$, $J_2$.

\textit{Modeling of off-Specular scattering using Distorted Wave Born Approximation formalism}: The simulated Off-Specular reflectivity profiles were generated by using Distorted-Wave Born-Approximation (DWBA) as implemented in the BornAgain~\cite{bornagain} software.\cite{Artur}. The basis for the model used in our simulation are based on the specular neutron reflectivity data at 300 K, 30 K, and 5 K (as shown in Fig. 2), fitting using Licorne-Py~\cite{licorne} software. In our model, we defined our sample as a multilayer consisting of the substrate with 3 layers (\textit{l}) on top, for which the scattering matrix elements can be expressed by:

\begin{equation*}
\braket{ \psi_i | \delta v | \psi_f } = \sum_l \sum_{\pm i} \sum_{\pm f} \braket{ \psi^\pm_{il} |\delta v | \psi^\pm_{fl}}
\end{equation*}
where, $\delta v$ is the first order perturbation expansion term of scattering length density $\left(v(\mathbf{r})\right)$, and $\psi_i$, $\psi_f$ denotes the incident, and final wavefunctions, respectively. The forward or backward traveling wavefunction in real-space is given by $\psi^+$  and $\psi^-$, respectively. The bottom-layer is composed of nanostructured silicon with honeycomb patterns, and the next layer, permalloy-layer, comprises of nanostructured permalloy. The top-layer is formed of a thin layer of partially oxidized permalloy. An ambient air layer was placed on top of the permalloy-layer. We have introduced a layout of honeycomb patterns of permalloy-hexagons with cylinders cut-out from the center, by using a hexagonal lattice with $a=31$ nm within the permalloy-layer. The form factor for the cylindrical cut-out is defined as: $F = 2\pi R^2 H \sinc \left( \dfrac{q_zH}{2} \right)\exp\left(\dfrac{iq_zH}{2} \right) \dfrac{J_1(q_{||}R)}{q_{||}R}$, where, $q_{||} \equiv \sqrt{q_x^2 + q_y^2}$, $J_1$ is a Bessel function of the first-kind, and radius, R = 11.2 nm, height, H = 13 nm. The magnetic phases were constructed by using rectangular elements with fixed magnetization directed along its length. The form factor of these rectangular elements is defined as: $F = LWH \sinc\left(\dfrac{q_xL}{2}\right)\sinc\left(\dfrac{q_yW}{2}\right)\sinc\left(\dfrac{q_zH}{2}\right)\exp\left(\dfrac{iq_zH}{2}\right)$, where $L$, $W$, and $H$ are 12.5 nm, 5 nm, and 13 nm, respectively. These magnetized-elements are placed as a part of the hexagonal lattice to incorporate long-range correlations with inter-cluster interference. The scattering matrix elements can be written as:
\begin{eqnarray*}
\braket{ \psi_i | \delta v | \psi_f } = \sum_j \exp\left({ i\mathbf{q}_{||}\mathbf{R}_{j||}}\right) \\
        \int d^2r_{||}\exp\left({i\mathbf{q}_{||}\mathbf{r}_{||}}\right) \int dz \phi^*_i(z) F(\mathbf{r} -\mathbf{R}_{j||}; \mathbf{T}_j)\phi_f(z) 
\end{eqnarray*}
where, $F(\mathbf{r}-\mathbf{R}_{ij}; \mathbf{T}_{j})$ is the form-factor for $\textbf{j}^{\text{th}}$ particle, such that $v_p(\mathbf{r}) = \sum_j F(\mathbf{r}-\mathbf{R}_{j||}; \mathbf{T}_j)$. The elastic scattering cross-section is given by: $\dfrac{d\sigma}{d\Omega} = |\braket{\psi_i | \delta v | \psi_f}|^2$. To account for the finite-size effect, we have used a 2D lattice interference function with a large isotropic 2D-Cauchy decay function with lateral structural correlation length of $\lambda_{x,y} = 1000$ nm. The position-correlation is given as: $\rho_S G(\mathbf{r}) = \sum_{m,n} \delta\left( \mathbf{r} - m\mathbf{a} -n\mathbf{b} \right) - \delta\left(\mathbf{r}\right)$, with lattice basis (\textbf{a, b}) and also introduced the effects of natural-disorder of the system by applying a small Debye-Waller factor corresponding to a position-variance of $\braket{\mathbf{x}}^2 = 0.1 $ nm$^2$. The interference function can be written as: 
\begin{equation*}
S({q}) = \rho_S \sum_{q_i \in \Lambda^*} \dfrac{2\pi\lambda_x\lambda_y}{\left(1+q_x^2\lambda_x^2 + q_y^2\lambda^2_y\right)^{{3}/{2}}}
\end{equation*}
For the simulation with just the structure, the magnetized elements were not introduced, and the sample contained only the coherent nuclear components. For the simulation with an external magnetic field, we allowed an external field commensurate with 0.5T and allowed higher magnetization in the magnetized-elements corresponding to the magnetic scattering length density obtained from specular reflectivity fitting. 

$\textit{Statistical Analysis}$: 1. Pre-processing of data: Experimental results are evaluated for statistical accuracy. For this purpose, neutron scattering measurements were performed for sufficiently longer amount of time to obtain statistically significant results. Experimental data is presented as it is, without any processing. 2. Data presentation: An error bar of one standard deviation is added in the data wherever applicable. It is mentioned in the figure caption of relevant figures. 3. Sample size: Experimental results were reproduced on two samples, fabricated under identical conditions. 4. Statistical methods: Neutron scattering method is a statistical probe. Measurements in different experimental conditions, e.g. different temperatures and fields, is used to reveal intrinsic ground state magnetic properties with sufficient details in magnetic honeycomb lattice. 5. Software used for statistical analysis: We have used LICORNE-PY software to analyze neutron reflectivity data. 

\section{Supporting Information}
Supporting Information is available from the Wiley Online Library or from the author.

\section{Acknowledgements}

We thank Artur Glavic, S. K. Kim and Giovanni Vignale for helpful discussion. The research at MU is supported by the U.S. Department of Energy, Office of Basic Energy Sciences under Grant No. DE-SC0014461. A portion of this research used resources at the Spallation Neutron Source, a DOE Office of Science User Facility operated by the Oak Ridge National Laboratory. 

\section{Conflict of Interest}
The authors declare no conflict of interest.

\section{Keywords}
Artificial magnetic honeycomb lattices, Geometric frustration, Magnetic charges, Degenerate states, Neutron reflectometry measurements.

\clearpage

\end{document}